\documentstyle[12pt]{article}
\newcommand{\eq}{\begin{equation}}
\newcommand{\en}{\end{equation}}
\newcommand{\eqa}{\begin{eqnarray}}
\newcommand{\ena}{\end{eqnarray}}
\newcommand{\eqs}{\begin{displaymath}}
\newcommand{\ens}{\end{displaymath}}
\newcommand{\eqas}{\begin{eqnarray*}}
\newcommand{\enas}{\end{eqnarray*}}

\advance\oddsidemargin by -\textwidth
\advance\oddsidemargin by -1in
\evensidemargin=\oddsidemargin
\begin{document}

$\mbox{ }$
\vspace{-3cm}
\begin{flushright}
\begin{tabular}{l}
{\bf KEK-TH-431 }\\
{\bf KEK preprint 95 }\\
March 1995
\end{tabular}
\end{flushright}

\baselineskip18pt
\vspace{1cm}
\begin{center}
\Large
{\baselineskip26pt \bf
A Background Independent Formulation of  \\
Noncritical String Theory
}
\end{center}
\vspace{1cm}
\begin{center}
\large
{\sc Nobuyuki Ishibashi and Hikaru Kawai}
\end{center}
\normalsize
\begin{center}
{\it KEK Theory Group, Tsukuba, Ibaraki 305, Japan}
\end{center}
\vspace{2cm}
\begin{center}
\normalsize
ABSTRACT
\end{center}
{\rightskip=2pc 
\leftskip=2pc 
\normalsize
Using the string field theory recently proposed by the authors
and collaborators, we give a background independent formulation of
rational noncritical string theories with $c\leq 1$.
With a little modification of the string field Hamiltonians previously
constructed,
we obtain string field theories which include various rational noncritical
string theories as classical backgrounds.
\vglue 0.6cm}

\newpage
\section{Introduction}
Constructing a field theory of string is one of the most important problems in
string theory\cite{SFT}.
One of the advantages of string field theories is that they
give a background independent formulation of string theory. In the first
quantized string theory, only perturbative analyses around a solution of
the classical equations of motion are possible.
The theory heavily depends on which
classical background is chosen. The problem is that innumerable
classical solutions are known now. In order to answer questions such as
which of the classical solutions (or which superposition of them)
describes the real world, we should have
a background independent formulation of string theory.

Recently a new class of string field theories of noncritical strings are
formulated by the authors and collaborators
\cite{IK}\cite{IK2}\cite{FIKN}\cite{IIKMNS}\cite{IIKMNS2}\cite{KKMW}.
Other recent developments are in \cite{others}.
It is natural to ask if such a formulation provides a
background independent one. The purpose of the present work is to answer this
question.

Let us briefly explain the formalism of our string field theory.
We introduce the creation (annihilation)
operator $\Psi^\dagger_i (l)$($\Psi_i (l)$) of a string. The argument $l$ here
denotes the length of the string and the subscript $i$ does the other degrees
of freedom. They
satisfy the following commutation relations:
\eq
\mbox{[} \Psi_i (l), \Psi_j^{\dagger }(l') \mbox{]}=\delta_{ij}\delta (l-l').
\label{comm}
\en
With the vacuum state $|0>$, satisfying $\Psi_i(l)|0>=<0|\Psi^\dagger_i(l)=0$,
one can express the $N$-string amplitude as
\eq
\lim_{D\rightarrow \infty}
<0|e^{-D{\cal H}}\Psi^\dagger_{i_1}(l_1)\cdots \Psi^\dagger_{i_N}(l_N)|0>.
\label{ampl}
\en
Here ${\cal H}$ is the Hamiltonian of the string field theory. The
Schwinger-Dyson (S-D)
equations for the string field are written as
\eq
\lim_{D\rightarrow \infty}\partial_D
<0|e^{-D{\cal H}}\Psi^\dagger_{i_1}(l_1)\cdots \Psi^\dagger_{i_N}(l_N)|0>=0.
\label{SD}
\en

The classical equations of motion of a field theory is nothing but the
classical
limit of the S-D equations. Therefore, we expect that eq.(\ref{SD}) with $N=1$
corresponds to the classical equation of motion of our string field theory,
when only the string tree level contributions are included. The classical
backgrounds are the solutions to this tree level S-D equation. We would like
to construct a Hamiltonian with which we can realize various first quantized
string theories as the solutions to the tree level S-D equation.

In section 2, we will construct the string field Hamiltonian which includes
the multicritical points of the one matrix model as the classical backgrounds.
In section 3, we will give a formulation in which $c\leq 1$ rational
string
theories are realized as the classical backgrounds. Section 4 is devoted to
the discussions.

\section{A Background Independent Formulation
of the $~~~$ Multicritical One Matrix Models}
The one matrix model has multicritical points\cite{kaz}, each of
which corresponds to a
$(2,2k-1)$ conformal field theory coupled to quantum gravity. These
multicritical points can be considered as the classical backgrounds of
one string theory. In this section, we would like to construct a string
field Hamiltonian which has these multicritical theories as classical
backgrounds.

Let us recall the Hamiltonian for the second critical point,
i.e. $c=0$ case\cite{IK}.
In this case, the only degrees of freedom of a string is its length and
we need no subscript $i$ for the string creation and annihilation operators
in eq.(\ref{comm}). The string field Hamiltonian becomes
\eqa
{\cal H}
&=&
\int_0^{\infty}dl_1\int_0^{\infty}dl_2
  \Psi^{\dagger}(l_1)\Psi^{\dagger}(l_2)\Psi (l_1+l_2)(l_1+l_2)
\nonumber
\\
& &
  +g\int_0^{\infty}dl_1\int_0^{\infty}dl_2
  \Psi^{\dagger}(l_1+l_2)\Psi (l_1)\Psi (l_2)l_1l_2
\nonumber
\\
& &
+\int _0^{\infty}dl\rho (l)\Psi (l).
\label{ham0}
\ena
The last term in the above is the tadpole term which describes the process
where a string with zero length disappears.
$\rho (l)$ is taken to be
$3\delta^{\prime \prime}-\frac{3}{4}t\delta(l)$ in which $t$ is the
cosmological
constant. The form of $\rho (l)$ is fixed to reproduce the disk amplitude for
$c=0$ string theory. The S-D equation eq.(\ref{SD}) for the disk amplitude
$f(l)$ is \cite{IK}
\eq
l\int_0^ldl'f(l')f(l-l')+\rho (l)=0,
\en
or
\eq
-\partial_{\zeta}(\tilde{f}(\zeta )^2)+\tilde{\rho}(\zeta )=0,
\label{lap}
\en
in the Laplace transformed form. Here
\eq
\tilde{f}(\zeta )=\int_0^\infty dle^{-\zeta l}f(l),~~
\tilde{\rho}(\zeta )=\int_0^\infty dle^{-\zeta l}\rho (l).
\en
In order for the $c=0$ disk amplitude
$\tilde{f}(\zeta )=(\zeta -\frac{1}{2}\sqrt{t})\sqrt{\zeta +\sqrt{t}}$
to be the solution for eq.(\ref{lap}), we should choose
\eq
\tilde{\rho}(\zeta )=3\zeta^2 -\frac{3}{4}t,
\en
or
\eq
\rho (l)=\delta^{\prime \prime}(l)-\frac{3}{4}t\delta(l).
\en

Now let us first consider how one can construct the string field Hamiltonian
corresponding to a multicritical point of the one matrix model.
In such a case, the disk
amplitude $\tilde{f}(\zeta )$ is of the form
\eq
\sqrt{Q(\zeta )},
\label{hdisk}
\en
where $Q(\zeta )$ is a polynomial of $\zeta$. $Q(\zeta )$ is of degree $2k-1$
for
the $k$-th critical point. Therefore, it is conceivable that one can obtain
the Hamiltonian by just changing the $\rho $ in eq.(\ref{ham0}) as
\eq
\tilde{\rho}(\zeta )=Q'(\zeta ).
\en

However this does not work. One can see that most easily in the following
manner. Let us decompose the Hamiltonian in eq.(\ref{ham0}) as
\eq
{\cal H}=\int_0^\infty dl [lT(l)+\rho (l)]\Psi (l),
\label{decom}
\en
where
\eq
T(l)=
\int_0^l dl'\Psi^\dagger (l')\Psi^\dagger (l-l')
+g\int_0^\infty dl'\Psi^\dagger (l+l')\Psi (l')l'.
\en
Then the S-D equation for the string field is equivalent to\cite{IK2}
\footnote{Notice that a kind of coherent state representation of $T(l)$ was
used in \cite{IK2}.}
\eq
<v|[lT(l)+\rho (l)]=0,
\label{SD2}
\en
with
\eq
<v|=\lim_{D\rightarrow \infty}<0|e^{-D{\cal H}}.
\en
In order for eq.(\ref{SD2}) to be integrable, the operator $lT(l)+\rho (l)$
should form a closed Lie algebra. In \cite{IK2} it was shown that this
condition
is related to a sort of the residual general coordinate invariance on the
worldsheet. The commutator of $lT(l)+\rho (l)$ was calculated to be
\eq
\mbox{[}l_1T(l_1)+\rho (l_1),l_2T(l_2)+\rho (l_2)\mbox{]}
=gl_1l_2\frac{l_1-l_2}{l_1+l_2}
\mbox{[}(l_1+l_2)T(l_1+l_2)+\rho (l_1+l_2)\mbox{]}
-gl_1l_2\frac{l_1-l_2}{l_1+l_2}\rho (l_1+l_2).
\label{lvir}
\en
The integrability requires that the last term on the right hand side
should vanish. In the Laplace transformed form, this means
\eq
\partial_{\zeta_1}\partial_{\zeta_2}(\partial_{\zeta_1}-\partial_{\zeta_2})
\frac{\int^{\zeta_1}_{\zeta_2}d\zeta '\tilde{\rho}(\zeta
')}{\zeta_1-\zeta_2}=0.
\label{alge}
\en
This equation is satisfied only when $\tilde{\rho}(\zeta )$ is a polynomial of
$\zeta$
with degree less than $4$. Then the algebra in eq.(\ref{lvir}) is equivalent
to the Virasoro algebra.
Therefore the S-D equation eq.(\ref{SD2}) is
consistent only for the first and the second critical points.

In order to express the higher critical points, we should change the form
of the Hamiltonian eq.(\ref{ham0}) without spoiling the Virasoro algebra.
The structures of the processes
corresponding to the splitting and the joining of the strings are very
firm and it is difficult to change them. The only modification one can think of
is to increase the power of $l$ in front of $T(l)$ in eq.(\ref{decom}):
\eq
{\cal H}=\int_0^\infty dl [l^nT(l)+\rho (l)]\Psi (l).
\label{mham}
\en
This modification means that when we consider the propagation of a string
with length $l$,
we take the time variable $D$ to be
$l^{n-1}\times (\mbox{geodesic~distance})$ instead of the geodesic distance
itself.
Then the S-D equation becomes
\eq
<v|[l^nT(l)+\rho (l)]=0.
\label{mSD}
\en
The disk amplitude of the $k$-th multicritical point is reproduced if
$\tilde{\rho}(\zeta )$ is a polynomial of degree $2k-1-n$. The integrability
condition is satisfied if
\eq
\partial^n_{\zeta_1}\partial^n_{\zeta_2}(\partial_{\zeta_1}-\partial_{\zeta_2})
\frac{\int^{\zeta_1}_{\zeta_2}d\zeta_1^\prime
\int^{\zeta_1^\prime}d\zeta_2^\prime \cdots
\int^{\zeta_{n-1}^\prime}d\zeta_n^\prime
\tilde{\rho}(\zeta_n^\prime )}{\zeta_1-\zeta_2}=0,
\en
namely if $\tilde{\rho}(\zeta )$ is a polynomial of degree less than
$n+3$.
Therefore if one chooses $n\geq k-1$, it seems that the Hamiltonian
in eq.(\ref{mham}) describes the $k$-critical points of the one matrix
model.

Indeed one can show that one can derive the Virasoro constraints\cite{FKN}
for the $k$-th critical point from the S-D equation eq.(\ref{mSD}) with
$\tilde{\rho}(\zeta )$ of degree $2k-1-n$ and $n\geq k-1$.
Let us define the generating functional of the
loop amplitudes as
\eq
Z(J)=<v|e^{-{\cal H}}e^{\int dlJ(l)\Psi^\dagger (l)}|0>.
\en
Eq.(\ref{mSD}) is rewritten in terms of $Z(J)$ as
\eq
[l^n\{ \int_o^ldl'\frac{\delta^2}{\delta J(l')\delta J(l-l')}
+g\int_0^\infty dl'l'J(l')\frac{\delta}{\delta J(l+l')}\}
+\rho (l)]Z(J)=0.
\label{rSD}
\en
This equation should be equivalent to the Virasoro constraints which
are relations between the correlation functions of the local operators
${\cal O}_r~r=n+\frac{1}{2},~n=0,1,2,\cdots$, with which the string creation
operator is expanded as\footnote{Here we take a convention different from
\cite{IK} but rather follow \cite{IIKMNS}. In the following, the indices $r,s$
should be understood as denoting half odd integers.}
\eq
\Psi^\dagger (l)=g\sum_{r>0}\frac{l^r}{\Gamma (r+1)}{\cal O}_r.
\en
In order to extract such informations, we should choose $J(l)$
so that
\eq
\int_0^\infty dlJ(l)l^r=g^{-1}\Gamma (r+1)x_r,
\en
where $x_r$ is the source for ${\cal O}_r$. Moreover, we should divide $Z(J)$
into the universal and the nonuniversal parts as $Z_{non}Z_{univ.}$ where
\eq
Z_{non}=\exp\{\int dlP(l)J(l)+\frac{1}{2}\int dldl'J(l)J(l')C_{non}(l,l')\}.
\en
$P(l)$ is the nonuniversal contribution from the disk amplitude, which is
of the form
\eq
\sum_{r>0}c_r\frac{l^{-r}}{\Gamma (-r+1)},
\en
for the higher critical points. $C_{non}(l,l')$ takes the same form for
all the critical points:
\eq
C_{non}(l,l')=
\frac{g}{2\pi}\frac{\sqrt{ll'}}{l+l'}.
\en
The universal part $Z_{univ.}$ should be considered as the generating function
of correlators:
\eq
Z_{univ.}(x)=<e^{\sum_rx_r{\cal O}_r}>.
\en

In the Laplace transformed form, eq.(\ref{rSD}) becomes the following equation
for $Z_{univ.}$:
\eqa
& &
[(-\partial_{\zeta})^n
\{\sum_{m=-1}^{\infty}\zeta^{-m-2}L_m+\sum_{r+s\geq 2}c_rc_s\zeta^{r+s-2}
\nonumber
\\
& &
\hspace{1cm}
+2g\sum_{r-s\geq 2}\zeta^{r-s-2}c_r\frac{\partial}{\partial x_s}
-g\sum_{r\geq \frac{5}{2}}c_r\sum_{j=1}^{r-\frac{3}{2}}\zeta^{r-j-\frac{3}{2}}
\int_0^\infty dl'J(l')\frac{l'^{-j+\frac{1}{2}}}{\Gamma (-j+\frac{1}{2})}\}
+\tilde{\rho}(\zeta )]Z_{univ.}=0.
\nonumber
\\
\label{vir}
\ena
Here $L_m$'s denote the Virasoro operators in \cite{FKN} and
\eqa
L_m
&=&
g\sum_{r+s=m}\frac{\partial^2}{\partial x_r^\prime \partial x_s^\prime}
+\sum_{s-r=m}r x_r^\prime \frac{\partial}{\partial x_s^\prime }
+\frac{1}{4g}\sum_{-r-s=m}rsx_r^\prime x_s^\prime +\frac{1}{16}\delta_{m,0},
\nonumber
\\
& &
\hspace{1cm}
x_r^\prime =x_r+\frac{2}{r}c_r,
\ena
in our notation.
In order for eq.(\ref{vir}) to be equivalent to the Virasoro constraints,
the terms except for the first term in eq.(\ref{vir}) should cancel with
each other.
This happens only when $c_r=0$ for $r\geq n+\frac{5}{2}$ and
\eq
(-\partial_\zeta )^n\sum_{r+s\geq 2}c_rc_s\zeta^{r+s-2}+\tilde{\rho}(\zeta )=0.
\en
Then the Virasoro constraints are derived from eq.(\ref{mSD}).

For the $k$-th critical points with the disk amplitude eq.(\ref{hdisk}) with
$\tilde{Q}(\zeta )$ a polynomial of  degree $2k-1$, $\tilde{P}(\zeta )$ is
of the form $c_{k+\frac{1}{2}}\zeta^{k-\frac{1}{2}}+\cdots$, with
$c_{k+\frac{1}{2}}\neq 0$.
We should choose $n\geq k-1$ and $\tilde{\rho}(\zeta )$ becomes a
polynomial of degree $2k-1-n$.

An interesting consequence of these considerations is the following. If one
takes $n>2k-1$, $\rho =0$. In this case, the disk amplitude $f(l)$ satisfies
the S-D equation
\eq
\partial_\zeta^n(\tilde{f}(\zeta )^2)=0.
\en
Although there are no tadpole terms, the disk amplitude can be nonzero because
of the integration constant of this equation. However, the disk amplitude for
the $k$-th critical point is not the only solution of this equation. An
amplitude of the form
\eq
\tilde{f}(\zeta )=\sqrt{Q(\zeta )},
\en
with $Q(\zeta )$ a polynomial of degree less than $n/2$ is also a solution.
Therefore we can choose $f(l)$ to be the disk amplitude for the $j$-th critical
point with $j<n/2$. Then the above arguments imply that all the string
amplitudes coincide with the amplitudes of the $j$-th critical point.
Hence the string field theory with the Hamiltonian in eq.(\ref{mham})
with $\rho =0$ has
the $j$-th critical point as a classical background. Therefore this Hamiltonian
gives a background independent formulation of $(2,2j-1)$ string theories
 with $j<n/2$.

We would like to conclude this section with the following remark
about the nature of the ``background'' we treated
in this section. Although we have a background independent formulation
of $(2,2j-1)$ string theories, the partition function depends on which
background is chosen\cite{DS}, even if it is calculated nonperturbatively.
As was discussed in \cite{SS}, these backgrounds are rather superselection
sectors. In our formalism, this fact manifests itself as follows.
For the Hamiltonian eq.(\ref{mham}) with $\rho =0$,
expression of the amplitudes in eq.(\ref{ampl}) does not have a definite
meaning if one does not choose the background. Indeed, with only
three string interaction terms, we can not expand it perturbatively. However
once one chooses a background with which one shifts $\Psi^\dagger$, one
obtains a Hamiltonian with a kinetic term and one can expand the amplitude
perturbatively. Therefore the expression eq.(\ref{ampl}) is
well-defined only after a background is chosen.

\section{A Background Independent Formulation of $c\leq 1$ String Theories}
We have shown that our string field theory provides a background independent
formulation of $(2,2k-1)$ string theories. In this section, we would like to
generalize this to the whole $(p,q)$ string theories. In the Liouville theory
approach, these theories can be considered as 2D string theories in
various dilaton and tachyon backgrounds. Therefore it is reasonable to
look for a string field theory which realizes these as classical backgrounds.

In order to do so, we will use the string field theories constructed
in \cite{IK2}. In that paper, we considered a string theory in which
the matter degrees of freedom are expressed by an RSOS-like lattice model.
Now the string fields $\Psi_i(l)$ have the subscript $i$ which
indicates a node of a Dynkin-like diagram. The Hamiltonian is
\eqa
{\cal H}
&=&
\sum_i\int_0^{\infty}dl_1\int_0^{\infty}dl_2
  \Psi^{\dagger}_i(l_1)\Psi^{\dagger}_i(l_2)\Psi_i (l_1+l_2)(l_1+l_2)
\nonumber
\\
& &
+\sum_{i,j}C_{ij}\int_0^{\infty}dl_1\int_0^{\infty}dl_2
  \Psi^{\dagger}_i(l_1+l_2)\Psi^{\dagger}_j(l_2)\Psi_i  (l_1)l_1
\nonumber
\\
& &
  +g\sum_i\int_0^{\infty}dl_1\int_0^{\infty}dl_2
  \Psi^{\dagger}_i(l_1+l_2)\Psi _i(l_1)\Psi _i(l_2)l_1l_2.
\label{dham}
\ena
Here, $C_{ij}$ is the connectivity matrix, where $C_{ij}=1$ when the heights
$i$ and $j$ are linked on the Dynkin-like diagram and it vanishes otherwise.
In \cite{IK2} we have shown that the S-D equation derived from this Hamiltonian
satisfies the integrability condition. The Hamiltonian can be decomposed as
\eq
{\cal H}=\sum_i\int_0^\infty dl lT_i(l)\Psi_i (l),
\en
and the operator $T_i(l)$ satisfies the decoupled Virasoro algebra:
\eq
[T_i(l_1),T_j(l_2)]=-g(l_1-l_2)T_i(l_1+l_2)\delta_{ij}.
\en
This is satisfied for any symmetric $C_{ij}$.
If one chooses $C_{ij}$ to
be the connectivity matrix of an $ADE$ or $\hat{A}\hat{D}\hat{E}$-type Dynkin
diagram, one can show that the disk amplitude of a $c\leq 1$ string theory
satisfies the tree level S-D equation\cite{IK2}\cite{kostov}.

The facts given above were the only justification in \cite{IK2} for taking
the Hamiltonian in eq.(\ref{dham}) as the one which describes the $c\leq 1$
string theories. Recently \cite{IIK} it is shown that the $W$ constraints can
be
derived from this Hamiltonian. Therefore now the Hamiltonian in eq.(\ref{dham})
is on firm ground.

We would like to obtain a background independent formulation of string
theories using this Hamiltonian. Following the previous section, let us
first modify the Hamiltonian as
\eq
{\cal H}=\sum_i\int_0^\infty dl l^nT_i(l)\Psi_i (l).
\label{dmham}
\en
The S-D equations for the disk amplitudes $f_i(l)$ become
\eq
\partial_\zeta^n[(\tilde{f}_i(\zeta ))^2+\sum_jC_{ij}
\int_C\frac{d\zeta ^\prime }{2\pi i}
\frac{\tilde{f}_i(\zeta^\prime )\tilde{f}_j(-\zeta^\prime )}
{\zeta -\zeta^\prime}
\mbox{]}=0.
\label{kostov}
\en
Here $C$ denotes the contour in the complex $\zeta '$ plane described in
Fig.1. In \cite{IK2}, we claimed that for $n=1$
these equations have a solution of the
form
\eq
\tilde{f}_i(\zeta )
=
v_i\mbox{[}
(\zeta +\sqrt{\zeta^2 -t})^\alpha +
(\zeta -\sqrt{\zeta^2 -t})^\alpha
\mbox{]},
\label{solf}
\en
where $\alpha$ and $v_i$ satisfy
\eq
\sum_j C_{ij}v_j=-2\cos (\pi \alpha )v_i.
\label{alv}
\en
If this is a solution for the $n=1$ case,
it is of course a solution for the $n>1$ case.
In order to prove that eq.(\ref{solf}) gives a solution to eq.(\ref{kostov}),
we should
perform the deformation of contour depicted in Fig.1. Strictly speaking, this
deformation is allowed if $n>2\alpha $. Therefore we will take $n$ to be
large enough here to make this deformation well-defined.

The disk amplitude for $(p,q)$ string theory has the form in eq.(\ref{solf})
with $\alpha =p/q$ or $q/p$\cite{MSS}. For example, if one takes the $A_{m-1}$
Dynkin diagram,
the values of $\alpha$ satisfying eq.(\ref{alv}) are
\eq
\frac{1}{m},~\frac{2}{m},~\cdots,\frac{m-1}{m}~~mod~1.
\en
Therefore, taking $n=3$, the string field Hamiltonian eq.(\ref{dmham}) for the
$A_{m-1}$ Dynkin diagram possesses the unitary $(m,m+1)$ string theory
as the classical background. Besides that, the disk amplitude for the
$(p,m),~(n=1,\cdots,m-1,m+2,\cdots ,[3m/2])$ string theories are also the
solutions to eq.(\ref{kostov}). Hence, with $n=3$ and taking various $A$-type
Dynkin diagrams, we can cover all the $(p,q)$ string theories. It is also
possible to take $D$, $E$ or even $\hat{A}\hat{D}\hat{E}$ Dynkin diagrams,
and cover such nondiagonal and $c=1$ models.

Of course, what we would like to do is not to construct the Hamiltonians for
various models, but to construct a Hamiltonian which includes these models
as classical backgrounds. In order to do so, we should notice the following
fact.
The S-D equations eq.(\ref{kostov}) for $A_{m-1}$ Dynkin diagram
have other special solutions. If the disk amplitudes
$f^\prime_i(l),~i=1,\cdots ,m'-1$ give a solution to
eq.(\ref{kostov}) for $A_{m'-1},~m'<m$ diagram, then it is trivial to see
that
\eqa
& &
f_i(l)=f_i^\prime (l),~i=1,\cdots ,m'-1,
\nonumber
\\
& &
f_i(l)=0,~i=m',\cdots ,m-1,
\ena
give a solution to the equation for the $A_{m-1}$ diagram.
It is in general true that a solution to the equations corresponding to
a subdiagram
\footnote{Here we define the notion ``subdiagram'' as a diagram whose nodes
form a subset of those of the original diagram and whose links are all the
links connecting these nodes in the original diagram. }
 gives a special solution to those corresponding to the original diagram.

This fact makes it possible for us to give a background independent
formulation of noncritical string theories. Indeed,
if one takes a big diagram, string field Hamiltonian corresponding to it
includes the string theories corresponding to the subdiagrams as
classical backgrounds. For example, taking the diagram as in Fig. 2, i.e.
$A_\infty$, one can obtain a string field theory which has all the
$A_{m}$ type models as classical backgrounds. We can go even further
 and taking the diagram as in Fig. 3, we can obtain a string field theory
which has all the rational $c\leq 1$ string theories corresponding to
$ADE$ and $\hat{A}\hat{D}\hat{E}$ Dynkin diagrams as classical
backgrounds.

Thus the Hamiltonian in eq.(\ref{dham}) corresponding to the diagram in
Fig. 3 is what we want.
We should remark that there can be many other classical backgrounds if
one takes the
diagram in Fig. 3. At present we are not sure about the nature of these
backgrounds and especially their relation to the $c=1$ barrier in two
dimensional quantum gravity\cite{KPZ}.

\section{Discussions}
In this work, we have presented string field Hamiltonians which
include various rational string theories as classical backgrounds.
It is shown that the Hamiltonian in eq.(\ref{mham}) with $\rho =0$ has
$(2,2j-1)~(j<n/2)$ string theories as classical backgrounds.
All the $ADE$ and $\hat{A}\hat{D}\hat{E}$ type string theories
are classical backgrounds of one string field theory described by the
Hamiltonian in eq.(\ref{dham}), where $C_{ij}$ is taken to be the connectivity
matrix of the big Dynkin diagram in Fig. 3.

We have not tried to construct
a field theory which includes irrational backgrounds. For example, $A_\infty$
diagram corresponds to a rational $c=1$ string theory. It can be considered as
a string theory on a discrete line. If one can take a ``continuum limit'' of
this theory, one may be able to obtain a irrational $c=1$ background.
Such a problem
is left as a future problem.

\section*{Acknowledgements}
We would like to thank M. Ikehara, T. Mogami and T. Yukawa for useful
discussions and comments.

\newpage
\section*{Figure Captions}
\begin{description}
\item{Fig. 1} The integration contour $C$ in eq.(\ref{kostov}) and
its deformation.
\item{Fig. 2} The $A_\infty$ Dynkin diagram.
\item{Fig. 3} The diagram which includes all the $ADE$ and
$\hat{A}\hat{D}\hat{E}$
Dynkin diagrams as subdiagrams.
\end{description}
\end{document}